\documentclass[9pt,twocolumn,twoside]{opticajnl}
\journal{opticajournal} 

\setboolean{shortarticle}{true}

\usepackage{lineno}
\usepackage{placeins}
\usepackage{comment}
\usepackage{subcaption}

\usepackage{pdfpages} 

\newcommand{\tc}[2]{#2}
\newcommand{\tcb}[1]{\tc{blue}{#1}}

\title{Photon Blockade Mediated by Two-Photon Absorption in an Optical Parametric Amplifier}

\author[1]{Weiyi An}
\author[2,3,*]{Jie Zhu}

\affil[1]{School of Optical and Electronic Information, Huazhong University of Science and Technology, 1037 Luoyu Rd., Wuhan, Hubei, 430074, China }
\affil[2]{Synopsys, Inc., 675 Almanor Ave, Sunnyvale, CA, 94085}
\affil[3]{Elmore Family School of Electrical and Computer Engineering, Purdue University, 610 Purdue Mall, West Lafayette, IN, 47907}

\affil[*]{zhujie@synopsys.com}

\begin{abstract}
Photon blockade (PB) is a quantum effect in strongly nonlinear systems where a single photon prevents the system from being excited to a higher level, generating anti-bunched light fields. It enables the generation of single-photon sources for quantum information processing. Conventional photon blockade (CPB) leverages strong nonlinear interactions to generate an anharmonic energy spectrum. Unconventional photon blockade (UPB) utilizes destructive quantum interference between excitation pathways. Recently, environmentally induced photon blockade (EPB) has emerged as a novel approach, exploiting two-photon absorption (TPA) to realize photon blockade. In this work, we combine UPB and EPB together, considering the TPA effect in the optical parametric amplifier (OPA), thereby achieving a more stable PB with stronger suppression of multi-photon states.
\end{abstract}

\setboolean{displaycopyright}{false} 

\begin{document}

\maketitle

With technological development in quantum computing and quantum communication, photon blockade (PB) is crucial for the realization of single-photon sources \cite{lounis2005single, eisaman2011invited} and quantum gates \cite{brylinski2002universal, barenco1995elementary}. In the current stage of research, the methods for achieving photon blockade can mainly be classified into three types: conventional photon blockade (CPB) \cite{Birnbaum2005, imamoḡlu1997strongly, leonski1994possibility, tian1992quantum}, unconventional photon blockade (UPB) \cite{snijders2018observation, flayac2017unconventional, Bamba2011, tang2015quantum, zhou2015unconventional, sarma2017quantum, sarma2018unconventional, lin2021realization}, and environmentally induced photon blockade (EPB) \cite{zhou2022environmentally, su2022nonlinear, zhang2023photon, feng2024photon}. CPB stems from the nonlinear splitting of the system's energy levels. When a single photon is excited, due to the detuning between the driving frequency and the transition frequency between the single-photon and multi-photon states, the energy level structure of the system repels the entry of the second photon, thereby achieving a stable output of the single-photon state. This usually requires strong nonlinearity and strong coupling of the system \cite{Birnbaum2005, imamoḡlu1997strongly, leonski1994possibility, tian1992quantum}\tc{blue}{, which led to a broad range of attempts to engineer stronger effective nonlinearities and coupling rates, for instance by optomechanical schemes \cite{gao2023phase-controlled} or optomagnonic microcavity \cite{gao2019photon}}. UPB is achieved through destructive interference between quantum paths. To suppress the appearance of the two-photon state. The requirement for system nonlinearity in UPB is relatively weaker \cite{snijders2018observation, flayac2017unconventional, Bamba2011, tang2015quantum, zhou2015unconventional, sarma2017quantum, sarma2018unconventional, lin2021realization}. EPB has been proposed recently. In these works, it is pointed out that the two-photon absorption (TPA) is conducive to achieving a stronger photon-blockade effect \cite{zhou2022environmentally, su2022nonlinear, zhang2023photon, feng2024photon}.

Optical parametric amplifier (OPA) is an optical amplification device based on nonlinear optical effects, achieving the amplification and frequency conversion of the signal light through the three-wave mixing process \cite{Baumgartner, Cerullo2003, manzoni2016design, wittmann2019taming, kuznetsov2025ultra, tzankov2003tunable,kumar2022low,dubietis2017ultrafast, sun2021optical,liu2010mid, kurdi2004optical}. In \cite{sarma2017quantum}, a model has been proposed to realize a strong
UPB by placing an OPA medium inside a Fabry-Perot cavity
under weak pump driving. The operation of OPA is often accompanied by TPA \cite{wittmann2019taming, tzankov2003tunable, manzoni2016design, kuznetsov2025ultra, kumar2022low, dubietis2017ultrafast,sun2021optical,liu2010mid, kurdi2004optical}. In the process of designing and manufacturing OPA, efforts are generally made to reduce problems such as gain limitation, nonlinear loss, thermal effect, and phase mismatch caused by TPA \cite{wittmann2019taming, tzankov2003tunable, manzoni2016design, kuznetsov2025ultra, kumar2022low, dubietis2017ultrafast, sun2021optical, liu2010mid, kurdi2004optical}.

In this work, we combine TPA and OPA to achieve a new photon blockade that has the advantages of both UPB and EPB. We first analytically derive the optimal operating condition using the effective non-Hermitian Hamiltonian. We then numerically calculate higher-order correlation functions to evaluate the effect of photon blockade. We demonstrat that the system of OPA with TPA has a stronger suppression of the two-photon state and simultaneously generates suppression of multi-photon states such as three-photon and four-photon states, which is not present without TPA. At the same time, the requirements of the system for parameters such as detuning and driving strength to achieve photon blockade will be relaxed, and the tolerance for fluctuations in actual parameters is higher.

\begin{figure}[tb]
\centering\includegraphics[width=0.35\textwidth]{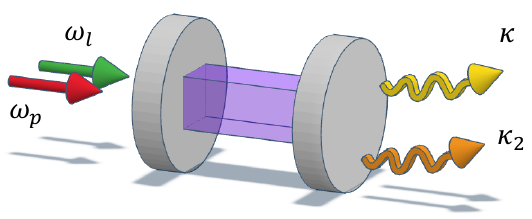}
\caption{The schematic of the system under our consideration. An OPA, pumped by a laser with a frequency of $\omega_p$, is coupled with a cavity. There is loss induced by single photon leakage and TPA, corresponding to dissipation rates $\kappa$ and $\kappa_2$ respectively.}
\label{fig:schematic}
\end{figure}

We consider a system of an OPA with a Fabry-Perot cavity, as shown in Fig. \ref{fig:schematic}. The system is driven by a laser at frequency $\omega_l$, and the OPA is pumped by another laser at frequency $\omega_p$, where $\omega_p=2\omega_a$. Under strong pumping conditions, the pumping laser can be regarded as a classical field. Setting $\hbar=1$, the Hamiltonian of this system can be written as (see supplemental document) \cite{Drummond_Hillery_2014, sarma2017quantum}
\begin{equation}
    H_0 = \omega_a a^{\dagger} a + ig(e^{i\theta} a^{\dagger 2} - e^{-i\theta} a^2) + \Omega(a^{\dagger} e^{i\omega_l t} + a e^{-i\omega_l t}),
    \label{eqs:original_Hamiltonian}
\end{equation}
where $a$ ($a^\dagger$) is the annihilation (creation) operator for the cavity mode, $\omega_a$ is the cavity-mode resonance frequency, and $\Omega$ is the intensity of the driving field. $g$ represents the nonlinear gain of the OPA. $\theta = \theta_0 + \omega_p t$, varying with time.

Performing a unitary transformation with $U = e^{i\omega_l t(a^\dagger a)}$, the Hamiltonian can be written in the rotating frame as
\begin{equation}
    H = \Delta_a a^{\dagger} a + ig(e^{i\theta_0} a^{\dagger 2} - e^{-i\theta_0} a^2) + \Omega(a^{\dagger} + a),
\end{equation}
where $\Delta_a=\omega_a-\omega_l$. We have used the working condition $\omega_p = 2\omega_a$ of the OPA to arrive at the aforementioned Hamiltonian.

Taking into account the dissipative effect of the environment, the evolution of the system can be described using the Lindblad master equation
\begin{equation}
    \frac{\partial \rho}{\partial t} = -i[H, \rho] + \mathcal{L}_\kappa\rho + \mathcal{L}_{\kappa_2}\rho,
    \label{eqs:master_equation}
\end{equation}
where $\mathcal{L}_\kappa\rho = \kappa (a\rho a^\dagger - \frac{1}{2}\{a^\dagger a, \rho\})$ is the Lindbladian that describes the dissipations of the cavity with a decay rate $\kappa$ and $\mathcal{L}_{\kappa_2}\rho = \kappa_2 (a^2\rho a^{\dagger 2} - \frac{1}{2}\{a^{\dagger2} a^2, \rho\})$ describes the TPA process with its decay rate $\kappa_2$.

To obtain approximate solution, the dissipative system can be described by an effective non-Hermitian Hamiltonian \cite{Plenio1998, feng2024photon, Wang2017, noh2020emission, Li:21}
\begin{equation}
    \label{eqs:effective_Hamiltonian}
    H^\prime = \Delta_a a^{\dagger} a + ig(e^{i\theta_0} a^{\dagger 2} - e^{-i\theta_0} a^2) + \Omega(a^{\dagger} + a)
    -i\frac{\kappa}{2}a^{\dagger} a-i\frac{\kappa_2}{2}a^{\dagger 2}a^2.
\end{equation}
In the weak-driving approximation, the Hilbert space can be expanded by Fock states with up to two photons (see supplemental document for the rationale of the approximation)
\begin{equation}
    \label{eqs:state_description}
    |\psi\rangle = C_{0}|0\rangle + C_{1}|1\rangle + C_{2}|2\rangle,
\end{equation}
where $C_0,C_1,C_2$ are the corresponding complex amplitudes.
Substituting Eq. (\ref{eqs:effective_Hamiltonian}) and (\ref{eqs:state_description}) into the Schrödinger equation $i\frac{d|\psi\rangle}{dt}=H^\prime|\psi\rangle$, the following set of differential equations are obtained:
\begin{equation}
\begin{aligned}
    &i\dot{C_0}=\Omega C_1-i\sqrt{2}ge^{-i\theta_0}C_2,\\
    &i\dot{C_1}=\Omega C_0+(\Delta_a-i\frac{\kappa}{2})C_1+\sqrt{2}\Omega C_2,\\
    &i\dot{C_2}=i\sqrt{2}ge^{i\theta_0}C_0+\sqrt{2}\Omega C_1 + 2(\Delta_a-i\frac{\kappa}{2})C_2-i\kappa_2 C_2.
\end{aligned}
\end{equation}
By using the perturbation method and considering the steady-state solutions \cite{huang2018nonreciprocal}, it is assumed that $C_0\approx1$ and $\dot{C_o}, \dot{C_1}, \dot{C_2}\approx0$. Therefore we have
\begin{equation}
\begin{aligned}
    (\Delta_a-i\frac{\kappa}{2})C_1+\sqrt{2}\Omega C_2 &= -\Omega,\\
    \sqrt{2}C_1+(2\Delta_a-i\kappa-i\kappa_2)C_2 &= -i\sqrt{2}ge^{i\theta_0}.
\end{aligned}
\end{equation}
Solving this two linear equations, we can obtain
\begin{equation}
\begin{aligned}
    \label{eqs:steady-state solution}
    &C_1 = \frac{\Omega[-2\Delta_a+i(2ge^{i\theta_0}+\kappa+\kappa_2)]}{(\Delta_a-i\kappa/2)[2\Delta_a-i(\kappa+\kappa_2)]-2\Omega^2},\\
    &C_2 = \frac{\sqrt{2}[ge^{i\theta_0}(-i\Delta_a -\kappa/2)+\Omega^2]}{(\Delta_a-i\kappa/2)[2\Delta_a-i(\kappa+\kappa_2)]-2\Omega^2}.
\end{aligned}
\end{equation}
From the above solutions, the optimum conditions of photon blockade, can be derived by letting $C_2=0$,
\begin{equation}
\begin{aligned}
    g_{opt}&=\frac{\Omega^2}{\sqrt{\Delta_a^2+\kappa^2/4}},\\
    \theta_{0opt}&=-\arctan(\frac{2\Delta_a}{\kappa}).
    \label{eqs:opt_conditions}
\end{aligned}
\end{equation}

The realization of PB can be analyzed by the second-order correlation function with zero time delay \cite{glauber1963coherent, glauber1963quantum,  titulaer1965correlation, mehta1965relation}. Under weak driving condition, we have
\begin{equation}
    \langle n\rangle=|C_1|^2+2|C_2|^2,
\end{equation}
\begin{equation}
    \label{eqs:g2(0)_expression}
    g^{(2)}(0)=\frac{2|C_2|^2}{(|C_1|^2+2|C_2|^2)^2}.
\end{equation}

\begin{figure}[tb]
\centering
\includegraphics[width=0.45\textwidth]{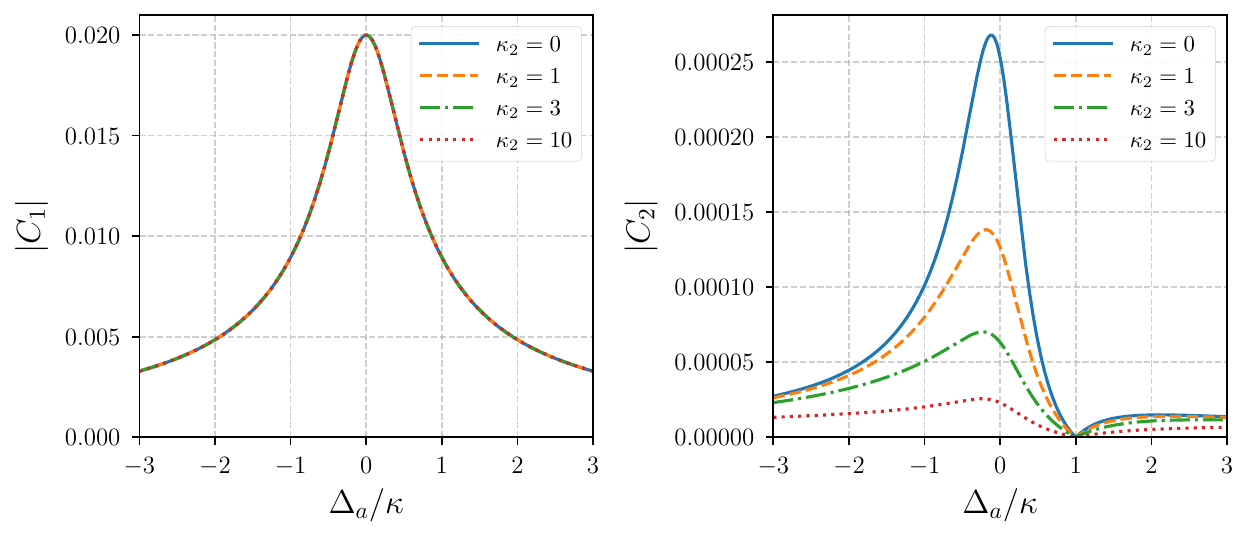}
\caption{The theoretical results of the amplitude of the steady state using the expression of Eq. (\ref{eqs:state_description}). $|C_1|$ and $|C_2|$ are shown in the left and right correspondingly.}
\label{fig:theoretical_amplitude}
\end{figure}

In Fig. \ref{fig:theoretical_amplitude}, we plot $|C_1|$ and $|C_2|$ according to Eq. (\ref{eqs:steady-state solution}).
Here we choose $\Omega/\kappa=0.01$, $g/\kappa=8.944\times10^{-5}$ and $\theta_0=-1.107\,\text{rad}$, such that the optimal result is obtained at $\Delta_a/\kappa=1$ \tc{blue}{according to Eq. (\ref{eqs:opt_conditions})}.
Theoretical calculations show that $|C_1|$ is hardly affected by TPA, while $|C_2|$ is \tc{blue}{significantly suppressed by the presence of two-photon absorption}. 

Next, we solve the master equation numerically under various conditions, and calculate the $n$-th correlation function
\begin{equation}
    g^{(n)}(0)=\frac{\text{Tr}(\rho_{ss}a^{\dagger n}a^n)}{[\text{Tr}(\rho_{ss}a^{\dagger}a)]^n},
    \label{eqs:ss_expression_of_gn}
\end{equation}
where $\rho_{ss}$ is the steady-state density matrix.
The numerical solution is obtained using QuTiP \cite{johansson2012qutip, JOHANSSON20131234, lambert2024qutip5quantumtoolbox} and MATLAB.
\begin{figure}[tb]
\centering
\includegraphics[width=0.4\textwidth]{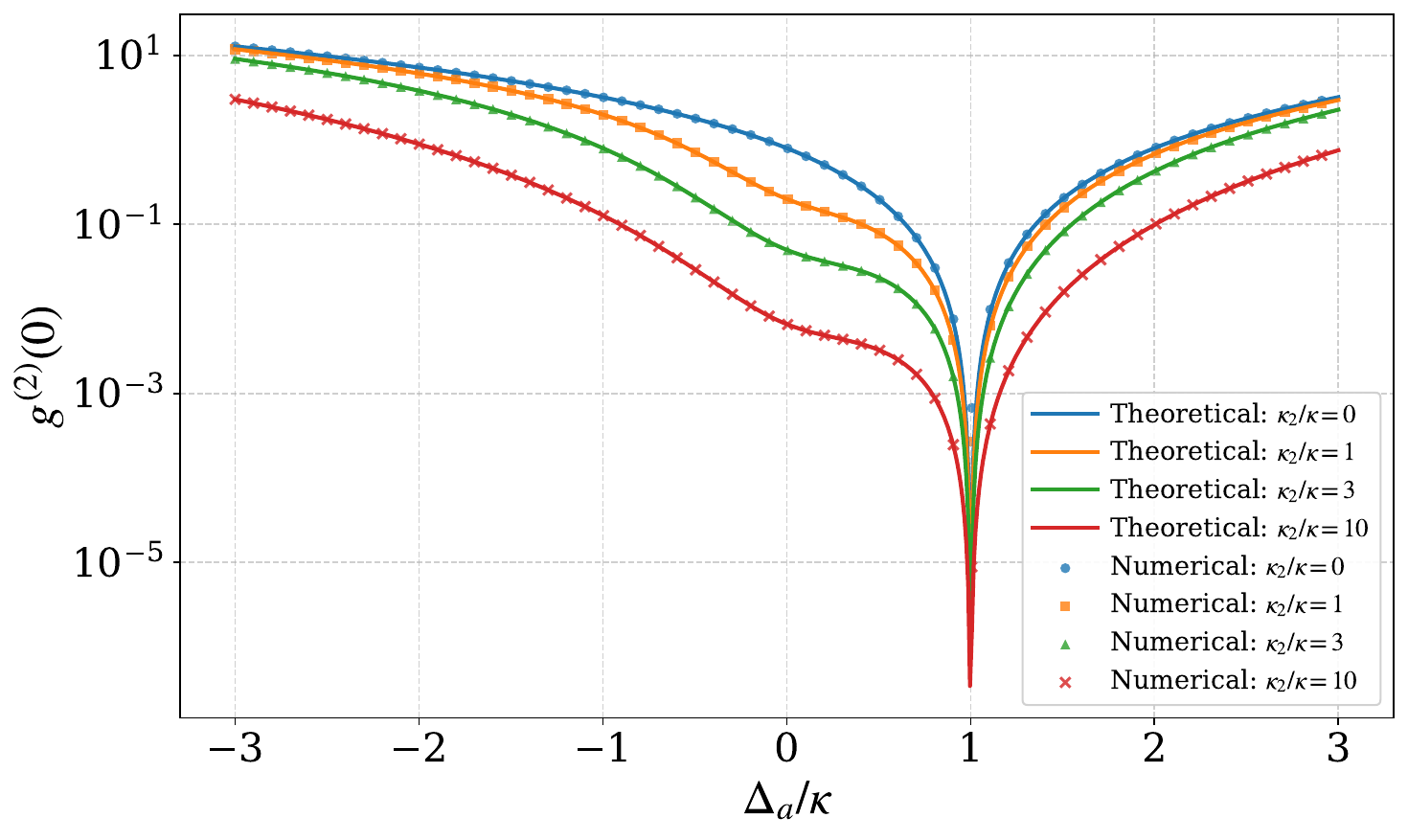}
\caption{The theoretical and numerical results of $g^{(2)}(0)$ \tc{blue}{obtained using the same parameters as in Fig. \ref{fig:theoretical_amplitude}, with the photon number truncated at 50 in the numerical calculations}.}
\label{fig:g2_theoretical&numerical}
\end{figure}
Fig. \ref{fig:g2_theoretical&numerical} shows the theoretical and numerical results under the conditions of different values of $\kappa_2$. The results of the numerical calculation agree well with the theoretical derivation. 

From another perspective, the previous work in Ref. \cite{sarma2017quantum} did not take into account higher-order correlation functions. In the previous work, due to considerations of weak driving, multi-photon states whose photon number is larger than two were ignored. However, the numerical solutions of the master equations indicate that when the system meets the optimal condition of photon blockade, the probabilities of detecting $|2\rangle$ and $|3\rangle$ get very close (see supplemental document) so that higher-order correlation functions should also be taken into account. 
\begin{figure}[tb]
    \captionsetup[subfigure]{position=below, justification=centering}
    \centering
    \subcaptionbox{$\kappa_2/\kappa=0$\label{subfig:gn_kappa_2=0}}[0.23\textwidth]{\includegraphics[width=0.9\linewidth]{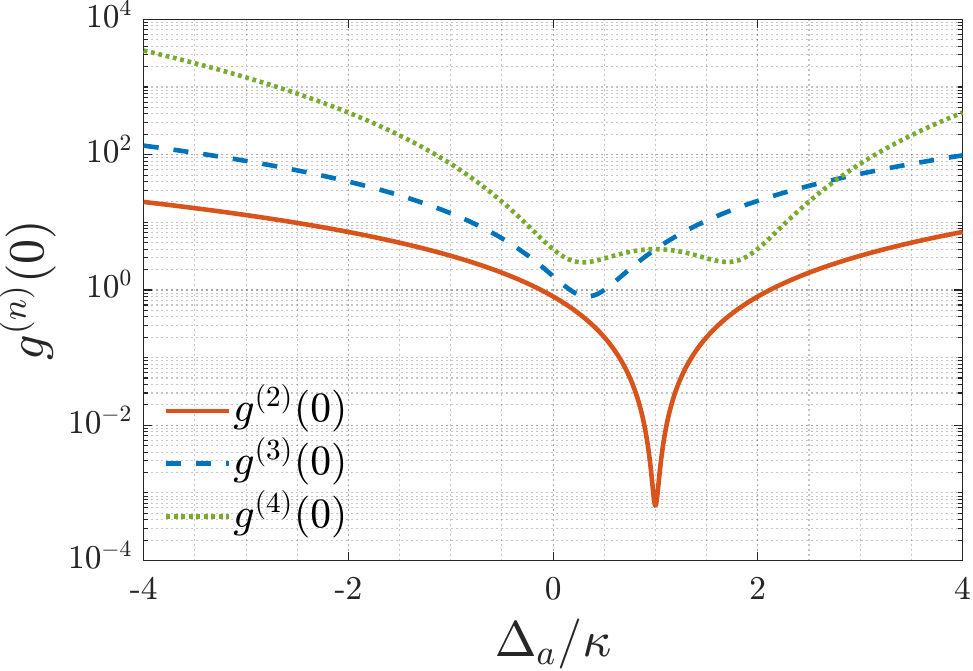}}\hfill
    \subcaptionbox{$\kappa_2/\kappa=1$}[0.23\textwidth]{\includegraphics[width=0.9\linewidth]{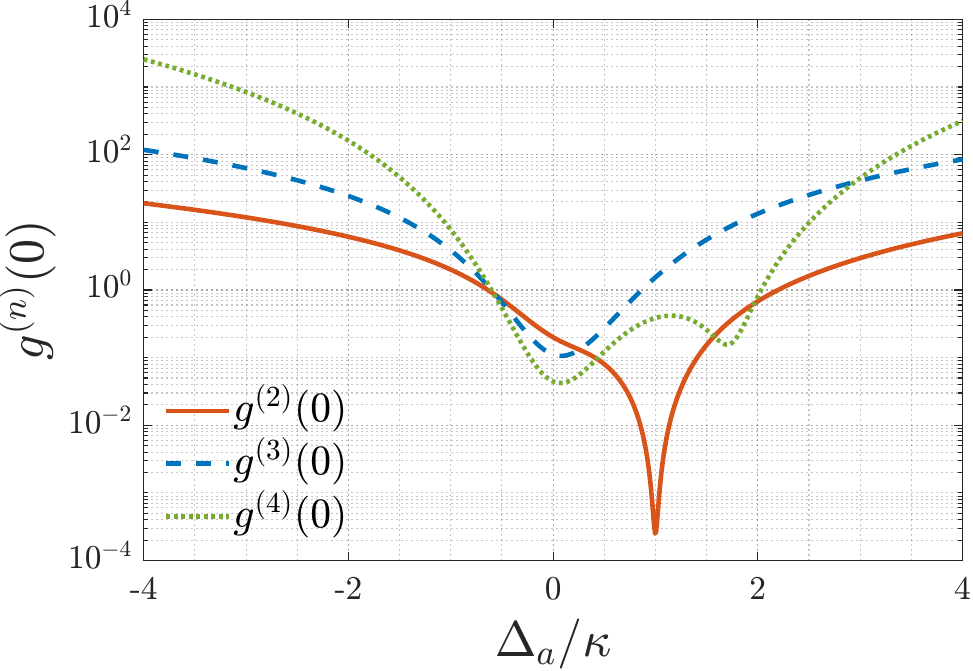}} \\
    \subcaptionbox{$\kappa_2/\kappa=3$}[0.23\textwidth]{\includegraphics[width=0.9\linewidth]{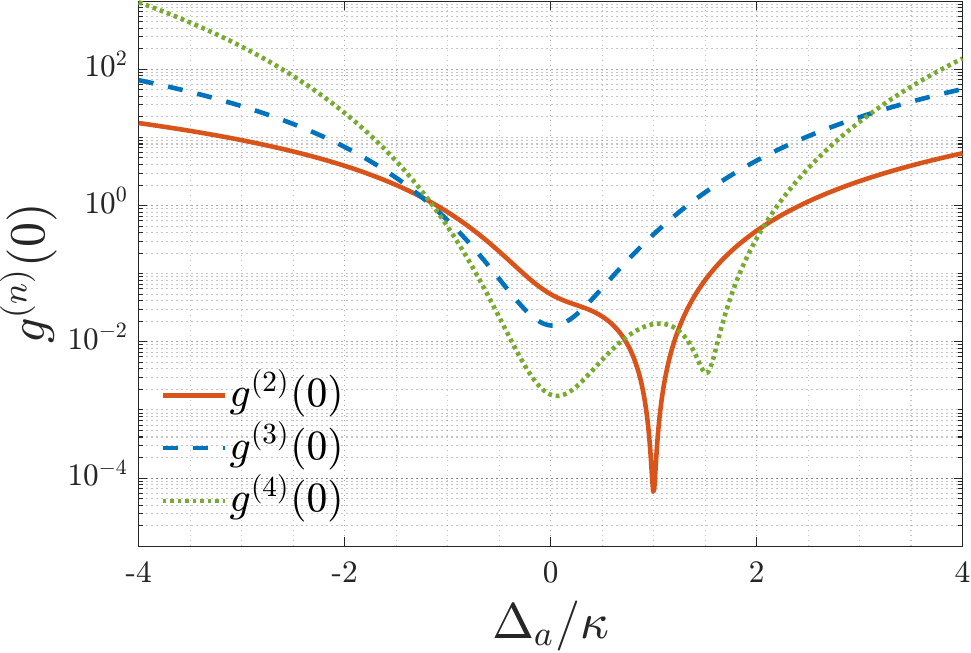}} \hfill
    \subcaptionbox{$\kappa_2/\kappa=10$}[0.23\textwidth]{\includegraphics[width=0.9\linewidth]{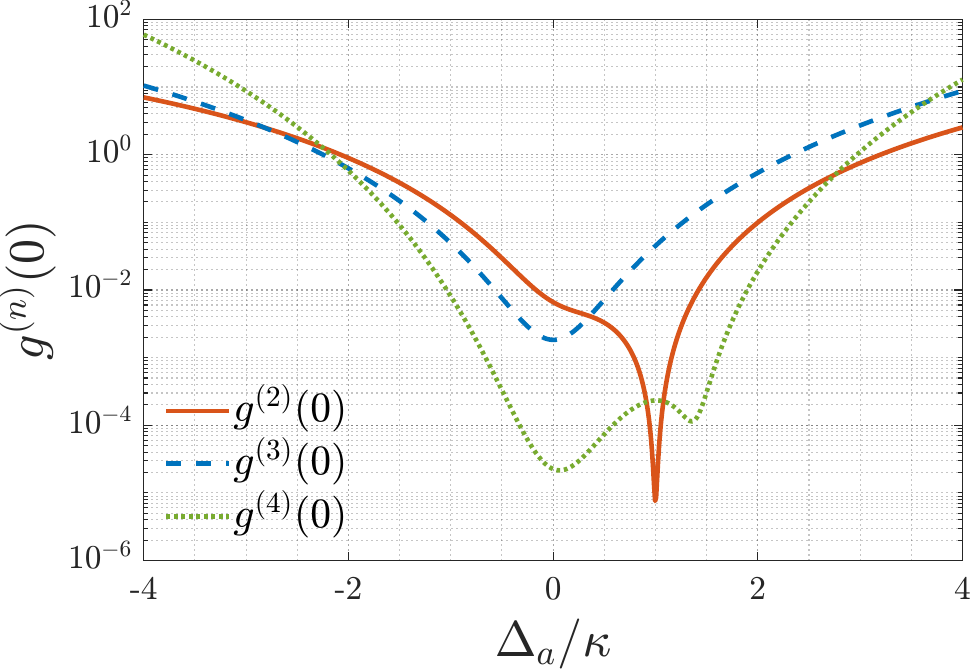}}
    \caption{$g^{(2)}(0)$, $g^{(3)}(0)$ and $g^{(4)}(0)$ under different strength of TPA \tc{blue}{obtained using the same parameters as in Fig. \ref{fig:theoretical_amplitude}, with the photon number truncated at 50 in the numerical calculations}.}
    \label{fig:gn(0)_of_different_kappa_2}
\end{figure}
As we can see in Fig. \ref{subfig:gn_kappa_2=0}, when TPA is not introduced, even if $g^{(2)}(0)$ takes a very small value near $\Delta_a/\kappa=1$, $g^{(3)}(0)$ and $g^{(4)}(0)$ are still greater than 1, indicating that the multi-photon entanglement still exists. This can be explained to some extent. When TPA is not introduced, UPB is achieved through interference between two quantum paths. Satisfying the destructive interference condition between the paths that lead to the two-photon state does not necessarily guarantee the destructive interference conditions between the paths that lead to other multi-photon states. Thus $g^{(3)}(0)$ and $g^{(4)}(0)$ can still reach a high value even if $g^{(2)}(0)$ is strong suppressed. After introducing TPA, $g^{(3)}(0)$ and $g^{(4)}(0)$ have decreased significantly. Moreover, the introduction of TPA has a negligible effect on the evolution of the second-order correlation function over time delay (see supplemental document). This indicates a better PB, as a result of the joint effect of UPB and EPB.

\begin{figure}[tb]
\centering\includegraphics[width=0.45\textwidth]{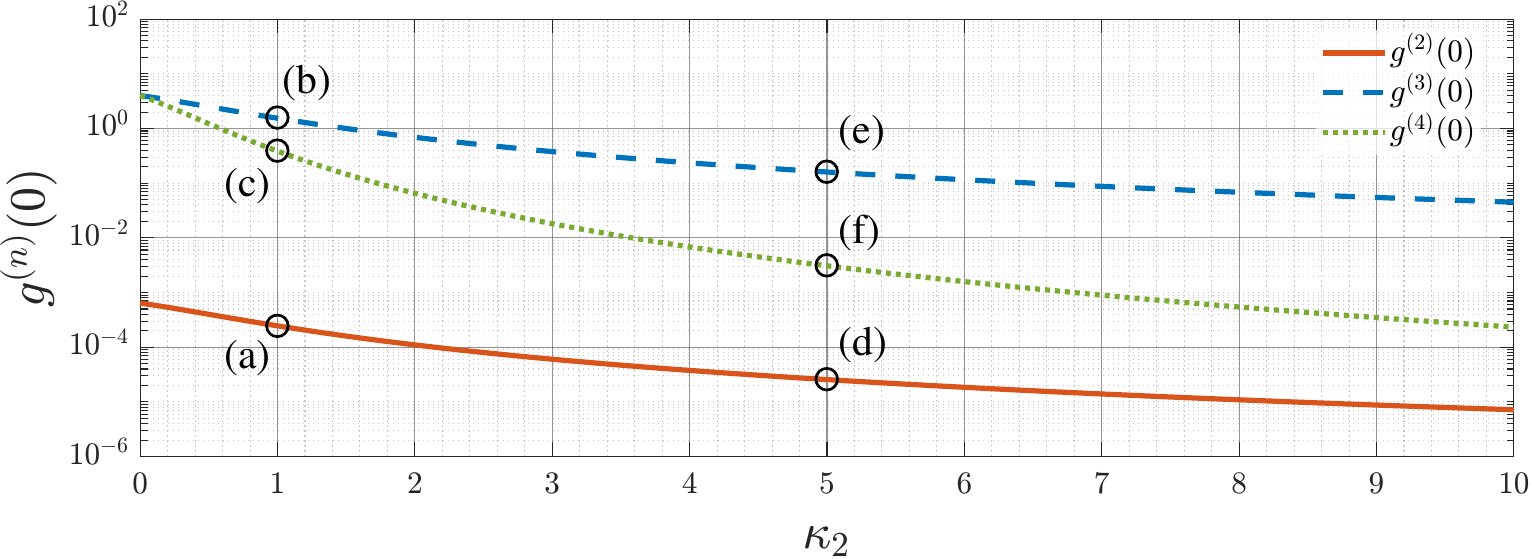}
\caption{\tc{blue}{$g^{(2)}(0)$, $g^{(3)}(0)$ and $g^{(4)}(0)$ under different TPA strengths at the optimal operating point as in Fig. \ref{fig:theoretical_amplitude}. Photon number is truncated at 50 in the numerical calculations. The marks (a)-(f) correspond to the subfigures in Fig. \ref{fig:six_points}. }}
\label{fig:optimal_points}
\end{figure}

\tcb{
To have a better understanding of the mechanism of the proposed PB at different $\kappa_2$, in Fig. \ref{fig:optimal_points}, we plot $g^{(n)}(0)$ as a function of $\kappa_2$ at the optimal $g$, $\theta_0$, and $\Omega$, for $\Delta_a/\kappa=1$. 
When $\kappa_2=0$, the system can be classified as a UPB as exemplified by $g^{(2)}(0)<1$ and $g^{(3)}(0),\,g^{(4)}(0)>1$ \cite{zhang2023photon}. In this regime, the PB effect arises from quantum interference.
As $\kappa_2$ is turned on, TPA gradually takes over. At $\kappa_2/\kappa\approx 2$, $g^{(3)}(0)$ and $g^{(4)}(0)$ both decrease to below 1 and the system lost the signature of UPB.
}

\begin{figure}[t]
\centering
\begin{subfigure}[b]{0.47\textwidth}
    \centering
    \includegraphics[width=\textwidth]{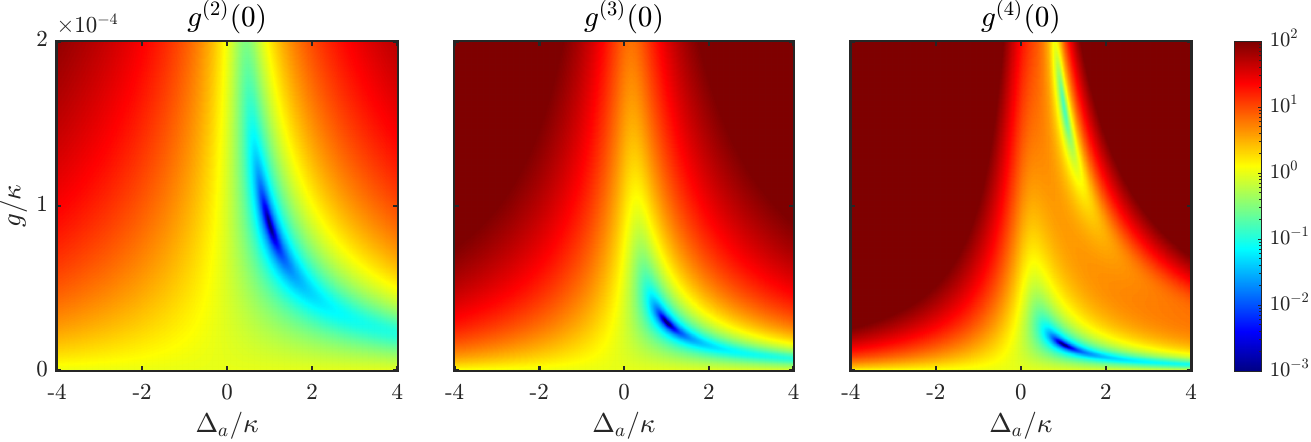}
    \caption{Heatmap of $g^{(n)}(0)$ under different values of $g$ and $\Delta_a$\tc{blue}{, with $\theta_0=-1.107\ \mathrm{rad}$ and $\Omega=0.01\kappa$}.}
    \label{fig:delta&g}
\end{subfigure}

\vspace{0.25cm}

\begin{subfigure}[b]{0.47\textwidth}
    \centering
    \includegraphics[width=\textwidth]{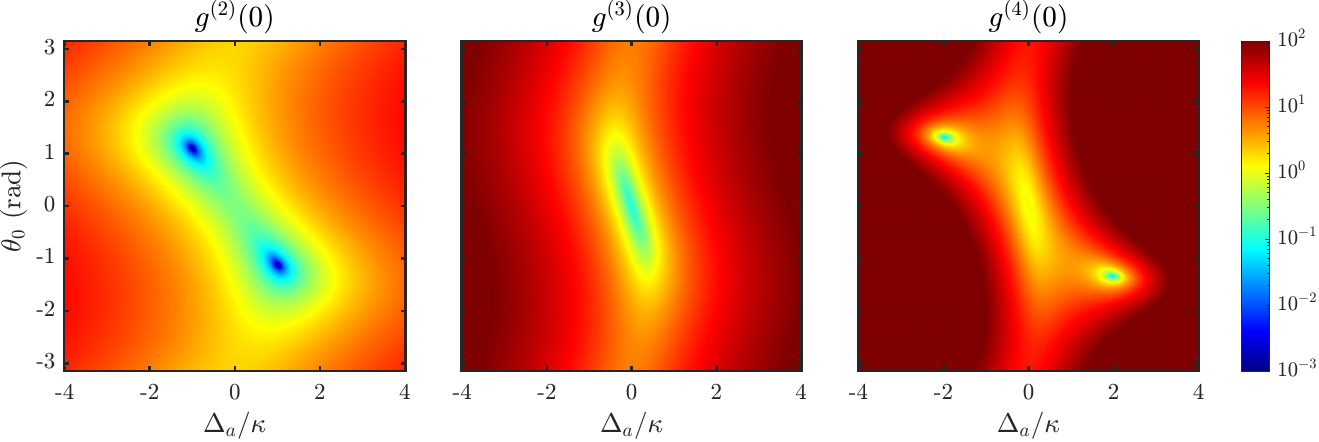}
    \caption{Heatmap of $g^{(n)}(0)$ under different values of $\theta_0$ and $\Delta_a$\tc{blue}{, with $\Omega=0.01\kappa$ and $g=8.944\times10^{-5}\kappa$}.}
    \label{fig:delta&theta}
\end{subfigure}

\vspace{0.25cm}

\begin{subfigure}[b]{0.47\textwidth}
    \centering
    \includegraphics[width=\textwidth]{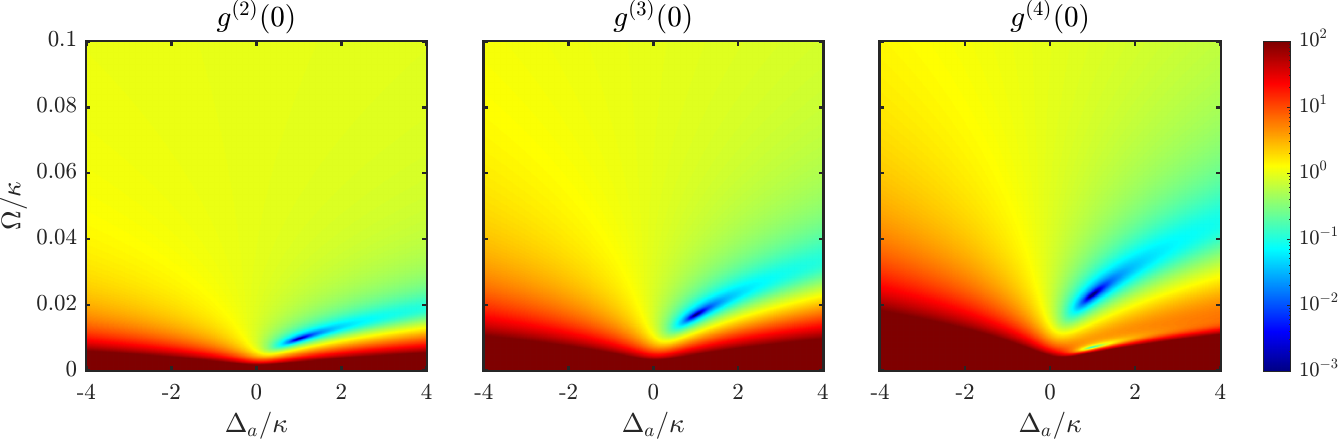}
    \caption{Heatmap of $g^{(n)}(0)$ under different values of $\Omega$ and $\Delta_a$\tc{blue}{, with $g=8.944\times10^{-5}\kappa$ and $\theta_0=-1.107\ \mathrm{rad}$}.}
    \label{fig:delta&Omega}
\end{subfigure}
\caption{Heatmaps of $g^{(n)}(0)$ under different parameters when $\kappa_2=0$ \tc{blue}{with the photon number truncated at 10 in the numerical calculations}.}
\label{fig:combined_2d}
\end{figure}

We then consider the impact of the fluctuation of $g$, $\theta_0$, $\Omega$, and $\Delta_a$ on the performance of the PB. In Fig. \ref{fig:combined_2d}, we first show the values of $g^{(n)}(0)$ under different parameters without TPA. 
\tcb{As expected, the minimal values of $g^{(n)}(0)$ are achieved at the optimal parameters according to Eq. (\ref{eqs:opt_conditions}).}
To measure the system’s sensitivity to parameter fluctuations at the onset of photon blockade, we define ``effective area'' to be the area where $g^{(n)}(0)$ is below 0.1, as the parameters of interest fluctuate. 
Fig. \ref{fig:delta&g} shows $g^{(n)}(0)$ as a function of nonlinear gain $g$ and detuning $\Delta_a$. Effective area decreases as $n$ increases.
Fig. \ref{fig:delta&theta} indicates that the correlation functions are very sensitive to phase mismatch. 
The variation of the driving intensity will also affect the performance of photon blockade, as shown in Fig. \ref{fig:delta&Omega}. However, unlike the influence of nonlinear gain and phase mismatch, effective area as a function of driving intensity and detuning increases when $n$ gets larger. 
This suggests that a small correlation function value at the optimal condition does not necessarily indicate a large effective area. Therefore, a good photon blockade design should take both into consideration.

\begin{figure}[tb]
\centering\includegraphics[width=0.47\textwidth]{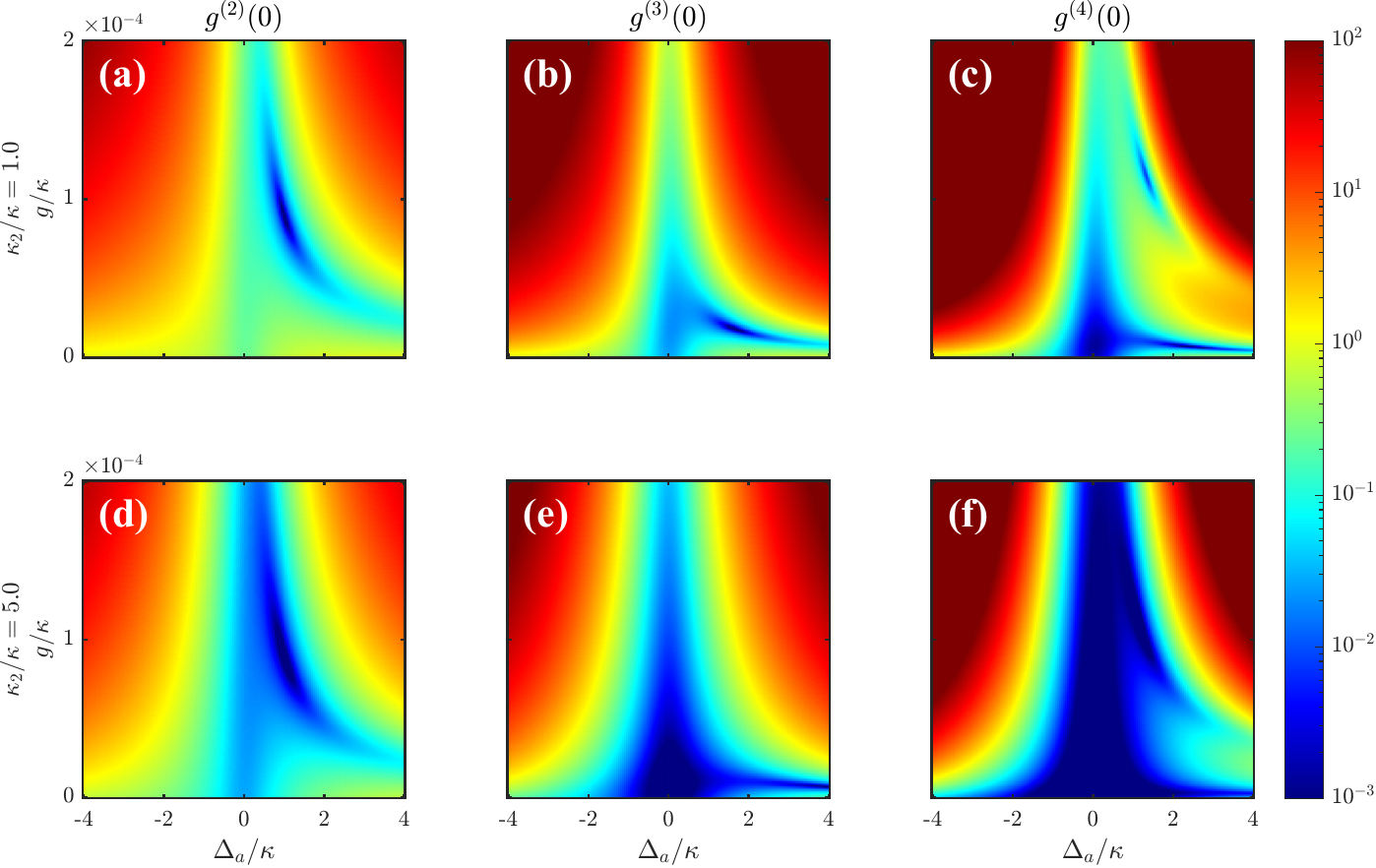}
\caption{\tcb{Heatmap of $g^{(n)}(0)$ under different values of $g$ and $\Delta_a$, with $\theta_0=-1.107\ \mathrm{rad}$ and $\Omega=0.01\kappa$. (a)-(c) $\kappa_2/\kappa=1.0$. (d)-(f) $\kappa_2/\kappa=5.0$. }}
\label{fig:six_points}
\end{figure}

The introduction of TPA can significantly increase the robustness of the system as indicated by the increase of effective areas. 
\tcb{
In Fig. \ref{fig:six_points}, we show $g^{(n)}(0)$ as a function of $g$ and $\Delta_a$, at two $\kappa_2$ values as marked in Fig. \ref{fig:optimal_points}. Compared with Fig. \ref{fig:delta&g}, the effective areas grow in both $\Delta_a$ direction and $g$ direction, suggesting that the PB effect relies less on quantum interference.
See supplemental document for the full parameter fluctuation analysis.
}

\tc{blue}{
The optimal parameters proposed in this work are achievable in cavity QED experiments . A typical PB setup \cite{hamsen2017two} with Fabry-Perot resonator reported a single-photon decay rate $\kappa/2\pi=2\,\mathrm{MHz}$ and a cavity driving strength comfortably approaching $\kappa$. A higher decay rate can be achieved in photonic crystal cavities. Large $\kappa_2$ can be obtained by III-V materials such as GaAs, GaN, and AlN \cite{gerace2014unconventional}.
The nonlinear gain $g$ of the OPA is proportional to the amplitude of the pump field driving the OPA and can be controlled flexibly.
}

In conclusion, we note that the TPA effect, commonly present in OPA processes due to crystal properties, is typically not favored for the purpose of optical amplification. However, in a system coupling an OPA to an optical cavity, this effect can be leveraged to facilitate enhanced photon blockade. Our theoretical analysis and numerical results demonstrate that TPA not only further suppresses the probability of multi-photon states but also relaxes the system parameter requirements for achieving photon blockade. This increased tolerance to parameter variations and fluctuations enhances the robustness of the blockade effect. Strategically combining optical amplification with TPA may therefore yield higher stability and controllability for photon blockade in practical systems, even in the presence of minor perturbations in optical intensity, frequency, and coupling strength. The above discussion is limited to numerical simulations; the actual effectiveness must await practical validation. Nevertheless, it offers a new conceptual avenue for future practical design.

\begin{backmatter}
\bmsection{Disclosures} 
The authors declare no conflicts of interest.

\bmsection{Supplemental document}
See Supplement 1 for supporting content.

\end{backmatter}

\bibliography{sample}
\bibliographyfullrefs{sample}







\section*{Supplementary material}

## 1. DERIVATION OF EQ. (1)

We consider a system where an OPA is coupled to a cavity, as shown in Fig. 1. There is a driving field driving this system, and the OPA is pumped by another laser. The Hamiltonian of this system can be written as (setting $\hbar = 1$) [1]

$$H_0 = \omega_a a^\dagger a + i\frac{\chi}{2}(a_p a^{\dagger 2} - a_p^\dagger a^2) + \Omega(a^\dagger e^{i\omega_l t} + a e^{-i\omega_l t}), \tag{S1}$$

where $a$ ($a^\dagger$) is the annihilation (creation) operator for the optical mode that we consider, and $a_p$ and $a_p^\dagger$ are the ones for the pump laser of the OPA. $\chi$ is the nonlinearity of the OPA due to a $\chi^{(2)}$ nonlinear process [1]. $\omega_a$ is the cavity-mode resonance frequency, and $\Omega$ and $\omega_l$, respectively, are the intensity and frequency of the driving field. Under strong pumping conditions, the pumping laser can be regarded as a classical field, $a_p \approx \alpha_p e^{i\omega_p t}$ [1, 2]. Let $ge^{i\theta} = \frac{\chi}{2}\alpha_p e^{i\omega_p t}$ where $\theta = \theta_0 + \omega_p t$, we have

$$H_0 = \omega_a a^\dagger a + ig(e^{i\theta} a^{\dagger 2} - e^{-i\theta} a^2) + \Omega(a^\dagger e^{i\omega_l t} + a e^{-i\omega_l t}), \tag{S2}$$

that is Eq. (1).

## 2. NUMERICAL VERIFICATION OF THE APPROXIMATE RATIONALITY

By numerically solve the Lindblad master equation

$$\frac{\partial \rho}{\partial t} = -i[H, \rho] + \mathcal{L}_\kappa \rho + \mathcal{L}_{\kappa_2} \rho \tag{S3}$$

with the system Hamiltonian given by

$$H = \Delta_a a^\dagger a + ig(e^{i\theta_0} a^{\dagger 2} - e^{-i\theta_0} a^2) + \Omega(a^\dagger + a), \tag{S4}$$

the steady-state density matrix can be calculated, and from it, one can obtain the detection probabilities of different Fock states, $\langle n|\rho|n\rangle$. Fig. S1 shows the numerical results under different $\kappa_2$. It can be seen that when $\Delta_a$ does not meet the optimal condition, the prerequisite for the approximation adopted in the theoretical analysis (i.e., ignoring Fock states higher than $|2\rangle$) is satisfied, that is, the probability of detecting higher Fock states are much smaller than that of $|2\rangle$. When $\Delta_a$ meets the optimal conditions, the probability of detecting $|2\rangle$ and $|3\rangle$ become very close, to the extent that it may be necessary to consider third-order or even higher-order correlation functions.

## 3. IMPACT OF TWO-PHOTON ABSORPTION ON DELAYED SECOND-ORDER CORRELATION FUNCTION $g^{(2)}(\tau)$

As suggested in [3], a weak nonlinearity may result in fast oscillations of the second-order correlation function on a time scale smaller than the cavity lifetime. Thus the temporal resolution of the state-of-the-art photodetectors may not capture sub-Poissonian light. To verify the practicability of the proposed system, we calculate the second-order correlation function with a time delay $\tau$ between two photodetectors [4, 5]

$$g^{(2)}(\tau) = \frac{\langle a^\dagger(0)\, a^\dagger(\tau)\, a(\tau)\, a(0)\rangle}{\langle a^\dagger(0)\, a(0)\rangle^2}. \tag{S5}$$

As shown in Fig. S2, numerical calculations indicate that the introduction of two-photon absorption has only a minor effect on the temporal resolution of $g^{(2)}(\tau)$ at small time delays $\tau$, and exhibits a negligible influence on macroscopic scales.

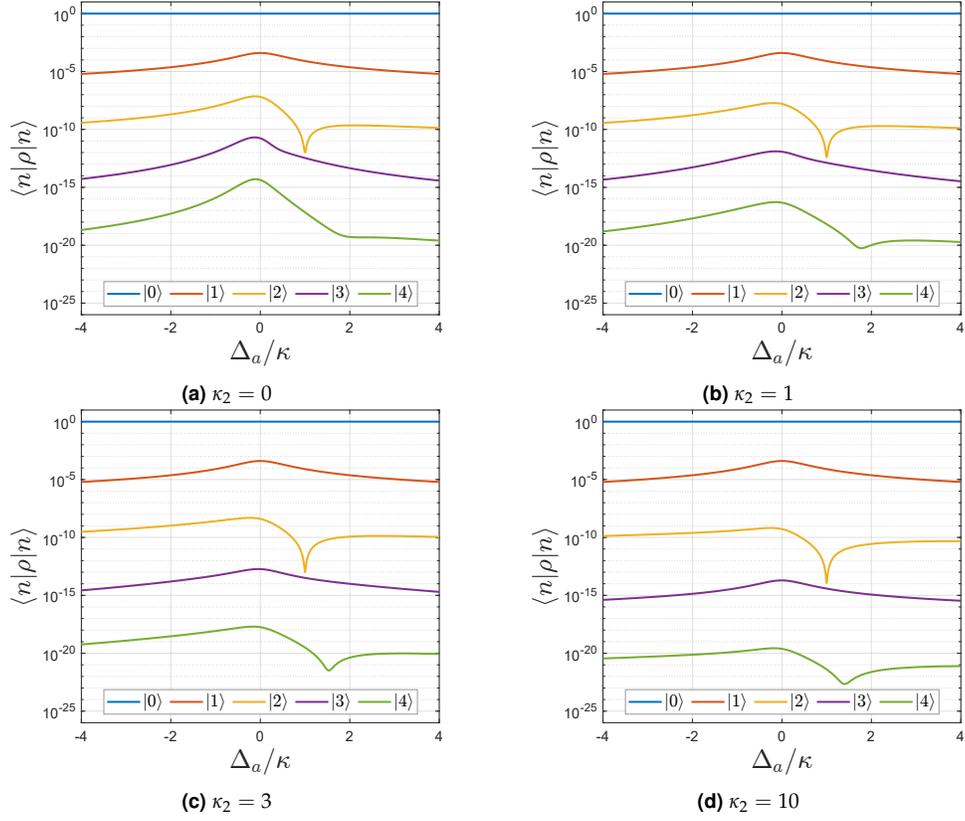

**Fig. S1.** Detection probabilites of Fock states under different strength of two-photon absorption.

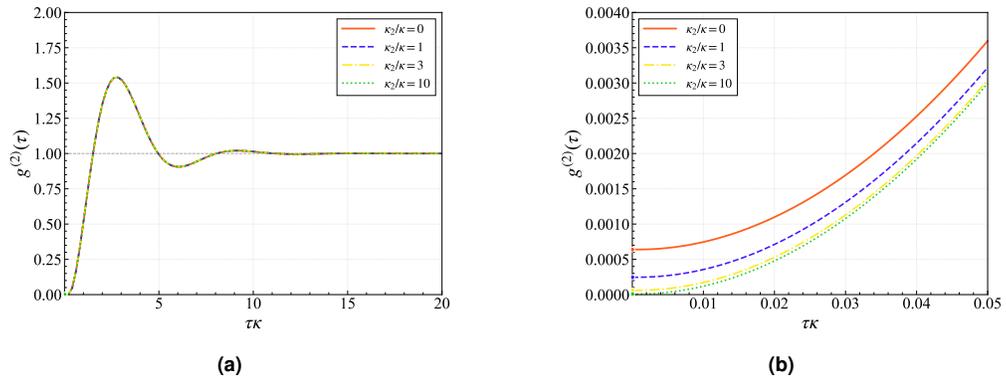

**Fig. S2.** Delayed second-order correlation function $g^{(2)}(\tau)$ in large (a) and small (b) ranges of $\tau\kappa$, at different TPA levels.

2



\end{document}